\documentclass[article]{emulateapj}

\newcommand{\sT}{\sigma_{\rm T}}

\newcommand{\e}{\epsilon}
\newcommand{\g}{\gamma}

\newcommand{\ep}{\epsilon^\prime}

\newcommand{\psim}{\lower.5ex\hbox{$\; \buildrel \propto \over\sim \;$}}
\newcommand{\lbar}{\lower.0ex\hbox{$\; \buildrel
{\lower0.0ex \hbox{-}} \over\lambda  \;$}}

\shorttitle{Constraints on the EBL}
\shortauthors{Finke \& Razzaque}

\begin{document}
\title {Constraints on the Extragalactic Background Light from 
Very High Energy Gamma-Ray Observations of Blazars}

\author{Justin D. Finke,$^{1}$ and Soebur Razzaque$^{1}$}

\affil{U.S.\ Naval Research Laboratory, Code 7653, 4555 Overlook SW,
       Washington, DC,
      20375-5352\\
$^1$NRL/NRC Research Associate}

\email{justin.finke@nrl.navy.mil}

\begin{abstract}

The extragalactic background light (EBL) from the infrared to the
ultraviolet is difficult to measure directly, but can be constrained
with a variety of methods.  EBL photons absorb $\g$-rays from distant
blazars, allowing one to use blazar spectra from atmospheric Cherenkov
telescopes (ACTs) to put upper limits on the EBL by assuming a blazar
source spectrum.  Here we apply a simple technique, similar to the one
developed by \citet{schroedter05_EBL}, to the most recent very-high
energy (VHE) $\g$-ray observations of blazars to put upper limits on
the EBL energy density.  This technique is independent of the EBL
model and has well-defined errors on the constraints.  Our results are
consistent with EBL measurements and constraints but marginally
inconsistent with several EBL models.

\end{abstract}


\section{Introduction}
\label{intro}

The EBL from the infrared through the visible and extending into the
ultraviolet is thought to be dominated by direct starlight emission
and absorption and re-radiation of starlight by dust.  The EBL photons
interact with VHE $\g$-rays from distant sources, such as blazars and
gamma-ray bursts, producing e$^+$-e$^-$ pairs and absorbing the VHE
photons.  Direct measurement of the EBL is difficult \citep[see][for a
review]{hauser01} due to contamination by foreground zodiacal and
Galactic light.  Galaxy counts may also be used to estimate the EBL,
but the unknown number of unresolved sources results in a lower limit.
Many models have been developed  
\citep{salamon98,malkan98,malkan01,stecker06,kneiske02,kneiske04,
primack01,primack05, primack08,gilmore08,franceschini08, razzaque09} 
but the input physics (e.g., star formation rate, dust absorption and
re-emission) are not constrained enough to give a reliable answer.

It is possible to use VHE $\g$-ray observations of blazars from ACTs
to constrain the EBL by assuming an intrinsic spectrum
\citep{stecker93,stanev98}. This was done by \citet{aharonian06} using
the hard observed spectrum of \object{1ES 1101-232} from HESS.  They
take a certain EBL shape \citep[that of ][]{aharonian03_1426} that is
consistent with EBL observations and lowered its normalization until
they can fit the computed intrinsic spectrum with a power law softer 
than $\Gamma_{int}=1.5$.  This gives an EBL quite close to the lower
limits from galaxy counts.  This technique has, however, been
criticized because it assumes a particular shape of the EBL spectrum,
which is not well known \citep{mazin07}.  This criticism applies to
several other constraints from blazar observations as well
\citep[e.g.,][]{aharonian07_0229}.  \citet{mazin07} have
addressed this problem by developing a sophisticated technique which
scans over a large grid of possible EBL shapes.  They de-absorb blazar
VHE $\g$-ray spectra with every possible shape, and throw out the
shapes which, when fit with a power-law, broken power-law, or
power-law with an exponential cutoff, give $\Gamma_{int}$ greater than
a certain value (they get two constraints, from $\Gamma_{int}>1.5$ and
$>0.75$).  The number of remaining allowed shapes give an upper limit
on the EBL intensity.  A similar technique was used by 
\citet{krennrich08} to rule out EBL models which are inconsistent 
with the recent lower limit at 3.6 $\mu$m \citep{levenson08}.  
A relatively simple technique was developed by
\citet{schroedter05_EBL} to give upper limits on the EBL which is
simpler than the technique of \citet{mazin07} and has well-defined
errors.  \citet{schroedter05_EBL} applied his technique to the 6
blazars seen with ACTs at that time to get upper limits on the EBL
energy density.  To date, VHE $\g$-rays have been seen from 21
blazars, including 1 flat spectrum radio quasar (FSRQ) and 20 
BL Lac objects.   In this paper, we apply the technique of
\citet{schroedter05_EBL} to an up-to-date sample of TeV blazar
spectra.

Note that this technique, and other techniques to constrain the EBL
with $\g$-ray observations from blazars do make assumptions about the
intrinsic spectrum of blazars, and is thus not a limit in the
conventional sense.  That is, understanding of blazars' intrinsic
spectra could improve in the future, which would change our
conclusions.  If blazars were found to typically have harder spectra
than we assume here, our upper limits would be weaker, whereas if they
were found to have typically softer spectra, they would be stronger.  

In \S\ 2 we describe how one can put upper limits on the $\g\g$
absorption optical depth, $\tau_{\g\g}(\e)$, and the EBL energy
density, $\e u_{EBL}(\e;z)$, using a method similar to the one
developed by \citet{schroedter05_EBL}.  In \S\ 3 we apply this
technique to the most recent ACT spectra from blazars and discuss our
results.

\section{Constraining the EBL energy density from Gamma-Ray Observations}
\label{EBLconstraint}

The observed $\nu F_{\nu}$ spectrum of a distant source, $f_{obs}$, at
redshift $z$ is related to its unknown intrinsic spectrum, $f_{int}$
by 
\begin{equation}
f_{obs}(\e) = e^{-\tau_{\g\g}(\e)}f_{int}(\e)\ ,
\end{equation}
where $\e$ is the dimensionless observed $\g$-ray photon energy.  The
greater the absorption, the harder the intrinsic spectrum must be to
produce a given observed spectrum.  The hardest spectrum observed in
the TeV range from a blazar near enough not to be strongly affected by
EBL attenuation is $\Gamma \approx 1.9$ (where $f(\e) =
f(\e_{min})(\e/\e_{min})^{-\Gamma+2}$) from \object{Mrk 501}
\citep{aharonian99_mrk501}.  Two blazars observed with EGRET have
$\Gamma \approx 1.5$, the FSRQ \object{0847-120} and the TeV-observed
high-peaked BL Lac \object{Mrk 501} \citep{nandikotkur07}.  The
hardest blazar with the Large Area Telescope (LAT) in the {\em Fermi
Gamma-Ray Space Telescope} 3-month bright AGN source list is
$\Gamma\sim 1.4$ from the BL Lac \object{MS 1050.7+4946} at $z=0.140$
\citep{abdo09_fermiAGN}.  This blazar has not been seen in the TeV
range, however, and may be a different class of blazar.  Just over
half of the TeV blazars have been seen with {\em Fermi} and all of the
TeV High-Peaked BL Lacs have {\em Fermi} spectral indices $\Gamma$
between 1.70 and 1.85.  From a theoretical standpoint,
shock-accelerated electrons are unlikely to produce $\gamma$-rays with
spectral indices harder than $\Gamma=1.5$ from Compton scattering.  
We will therefore conservatively take $\Gamma_{int} \ge
\Gamma^{min}_{int} = 1.5$ as the {\em conventional limit} on the
spectral index.  Several theoretical possibilities have been proposed
to create harder spectra.  \citet{katarzynski06} have suggested
that electrons create synchrotron and Compton emission at a
significant distance from the acceleration region, leading to a high
cutoff in the lower portion of the electron spectrum that generates
the TeV $\gamma$-rays.  This could result in harder observed spectra.
\citet{stecker07_accel} have performed simulations which generate
harder electron spectra at the highest energies in relativistic
shocks, and hence harder observed TeV $\gamma$-ray spectra.  Internal
$\g\g$ absorption could possibly produce harder spectra in the TeV
range \citep{aharonian08}, although this would probably apply
more to distant FSRQs with significant scattered radiation fields
(i.e., broad line regions) which we do not consider in this paper.
The decay of pions from a hadronic source could also produce an
extremely hard TeV component, independent of the lower energy electron
synchrotron emission \citep{muecke03}.  Compton scattering of the
cosmic microwave background radiation in the extended jets of blazars
could produce harder spectra \citep{boett08}.  These are
possibilities that must be considered.  Furthermore, creating a model
consistent with the recent galaxy count result at 3.6 $\mu$m by
\citet{levenson08}, \citet{krennrich08} find that spectral indices
harder than $\Gamma_{int} \approx 1.5$ seem necessary.  We will,
therefore, take $\Gamma_{int} \le \Gamma_{int}^{min} = 1.0$ as the
{\em extreme limit}.

By assuming the intrinsic spectrum is
limited by $\Gamma_{int}^{min}$ we can put an upper limit on
$\tau_{\g\g}(\e)$,
\begin{eqnarray}
\label{taumax}
\tau_{\g\g}^{max}(\e, z) = 
	\ln\left[\frac{f_{int}(\e_{min})}{f_{obs}(\e_{min})}\right] + 
\nonumber \\
	(\Gamma_{obs}-\Gamma_{int}^{min})\ln(\e/\e_{min})\ .
\end{eqnarray}
The standard error of $\tau_{\g\g}^{max}$ is given by
\begin{eqnarray}
\label{sigmataumax}
\sigma(\tau_{\g\g}^{max}(\e, z)) = 
\nonumber \\
	\sigma(f_{obs}(\e)) / f_{obs}(\e)\ .
\end{eqnarray}
We use the $\gamma$-ray flux at $E_{min} = \e_{min} m_e c^2$ TeV, the
lowest energy bin of the observation, to normalize the observed
spectra.  We then use the EBL model which gives the greatest
$\tau_{\g\g}(\e_{min})$ (which is almost always the fast evolution
model of \citet{stecker06}) to de-absorb the spectra here, i.e.,
$f_{int}(\e_{min}) = f_{abs}(\e_{min})e^{\tau_{\g\g}(\e_{min})}$, and
use this to normalize our intrinsic maximum spectra.

Once $\tau_{\g\g}^{max}(\e,z)$ is found, one can calculate an upper
limit on the local EBL.  This is done using several approximations, as
follows.  We begin with the $\g\g$ opacity of the universe as a
function of the comoving EBL energy density, which is given by
\begin{eqnarray}
\label{taugg}
\tau_{\g\g}(\e_1, z) = \frac{c\pi r_e^2}{m_ec^2 \e_1^2}\ 
\int^z_0 \frac{dz^\prime}{(1+z^\prime)^2}\ 
\left| \frac{dt_*}{dz^\prime}\right|\ 
\nonumber \\
\int^\infty_{\frac{1}{\e_1(1+z^\prime)}} d\ep\ 
\frac{u^\prime_{EBL}(\ep;z^\prime)}{\e^{\prime 3}}\ 
\bar{\phi}(\ep\e_1(1+z^\prime))
\end{eqnarray}
where, in a flat $\Lambda$CDM universe,  
\begin{eqnarray}
\label{dtdz}
\left| \frac{dt_*}{dz^\prime}\right|\ = \frac{1}
{H_0(1+z^\prime)\sqrt{\Omega_m(1+z^\prime)^3 + \Omega_\Lambda}}\ ,
\end{eqnarray}
$u^\prime_{EBL}(\ep;z^\prime)$ is the comoving EBL energy density per unit 
comoving dimensionless energy, 
$\ep=\e(1+z)$,  
\begin{eqnarray}
\bar{\phi}(s_0) = \frac{1+\beta_0^2}{1-\beta_0^2}\ln w_0 - 
\beta_0^2\ln w_0 - \frac{4\beta_0}{1 - \beta_0^2}
\nonumber \\ 
+ 2\beta_0 + 4\ln w_0 \ln(1+w_0) - 4L(w_0)\ ,
\end{eqnarray}
$\beta_0^2 = 1 - 1/s_0$, $w_0=(1+\beta_0)/(1-\beta_0)$, and 
\begin{eqnarray}
L(w_0) = \int^{w_0}_1 dw\ w^{-1}\ln(1+w)\ 
\end{eqnarray}
\citep{gould67,brown73}.  

If we assume the blazar is at low redshift, then $|dt_*/dz| \approx
H_0^{-1}$, where we use $H_0 = 70$ km s$^{-1}$ Mpc$^{-1}$ for the
Hubble constant.  We also use the approximation that the EBL does not
change significantly at low redshifts, and that all of the EBL photons
are at the energy where the pair-production cross-section is largest,
$\ep_*=2/(\e_1(1+z^\prime))\approx2/\e_1$ (i.e., we assume the
EBL is monochromatic).  The latter approximation overestimates the
actual EBL energy density, since in reality the absorption will be
spread out over a range of EBL photon energies, so that we are left
with an upper limit on the EBL energy density, $\ep u^{\prime\
max}_{EBL}(\e;z)$.  Using the Dirac delta-function these
approximations are written as
\begin{eqnarray}
\label{eblapprox}
u^\prime_{EBL}(\ep;z\approx0) \approx 
	\ep_* u^{\prime\ max}_{EBL}\left(\ep_*;z\approx 0 \right)\ 
	\delta\left(\ep - \ep_*\right)\ .
\end{eqnarray}
Eqn. (\ref{eblapprox}) allows us to perform the integrals in
eqn. (\ref{taugg}) so that 
\begin{eqnarray}
\label{tauggapprox}
\tau^{max}_{\g\g}(\e_1,z\approx0) = \frac{c\pi r_e^2\e_1}{8m_e c^2}\ 
\nonumber \\ \times
	\frac{\ep_* u^{\prime\ max}_{EBL}(\ep_*;z\approx0)}{H_0}\ 
\bar{\phi}(2)\ z \ 
\end{eqnarray}
or
\begin{eqnarray}
\label{nEBLmax}
\ep_* u_{EBL}^{\prime\ max}(\ep_*;z\approx0) = \frac{64m_ec^2H_0}
	{3c\sT z \bar{\phi}(2) \e_1}\ 
\nonumber \\ \times
	\tau_{\g\g}^{max}(2/\ep_*,z\approx0)
\end{eqnarray}
where $\bar{\phi}(2) \approx 1.787$.  The error in this EBL limit 
is given by
\begin{eqnarray}
\sigma(\ep_*u^{\prime max}_{EBL}) = 
\frac{64m_ec^2H_0}{3c\sT \bar{\phi}(2) z \e_1}\ 
	\sigma(\tau_{\g\g}^{max})
\end{eqnarray}
where $\sigma(\tau_{\g\g}^{max})$, the error in $\tau_{\g\g}^{max}$, 
is given by eqn. (\ref{sigmataumax}).  
For $z\approx0$, the comoving EBL energy density  
will be equal to the observed energy density, i.e., 
\begin{equation}
\e_*u_{EBL}(\e_*,z\approx0)=\ep_*u^\prime_{EBL}(\ep_*,z\approx0)\ .  
\end{equation}
For higher redshifts, 
\begin{eqnarray}
\e_* u_{EBL}(\e_*;z) = (1+z)^{-4} \ep_*\ u^\prime_{EBL}(\ep_*;z)\ .  
\end{eqnarray}
The EBL energy density can be converted to intensity 
in units of, e.g., nW m$^{-2}$ sr$^{-1}$ by 
\begin{eqnarray}
\e I_\e(z) = \frac{c}{4\pi}\ \e_* u_{EBL}(\e_*;z)\ .
\end{eqnarray}

By inspecting eqns. (\ref{taumax}) and (\ref{nEBLmax}) we 
see that $\e u_{EBL}^{max}(\e;z)$ will be minimized, thus giving 
the strongest constraint, if the dimensionless parameter 
\begin{eqnarray}
\xi \equiv \left(\frac{TeV}{E_{max}}\right)
	\frac{\Gamma_{obs}-\Gamma_{int}^{min}}{z}
	\ln\left(\frac{E_{max}}{E_{min}}\right)
\end{eqnarray}
is minimized, where $E_{max}=m_ec^2\e_{max}$ is the energy of the
highest energy photon bin observed from a source.   This
parameter should be seen as a general guide.  The highest energy
photon bin is often not well observed, for example, in \object{H
1426+428} the highest energy bin is less than a $2 \sigma$ detection
\citep{aharonian03_1426}.  

\begin{figure*}
\epsscale{1.0}
\plotone{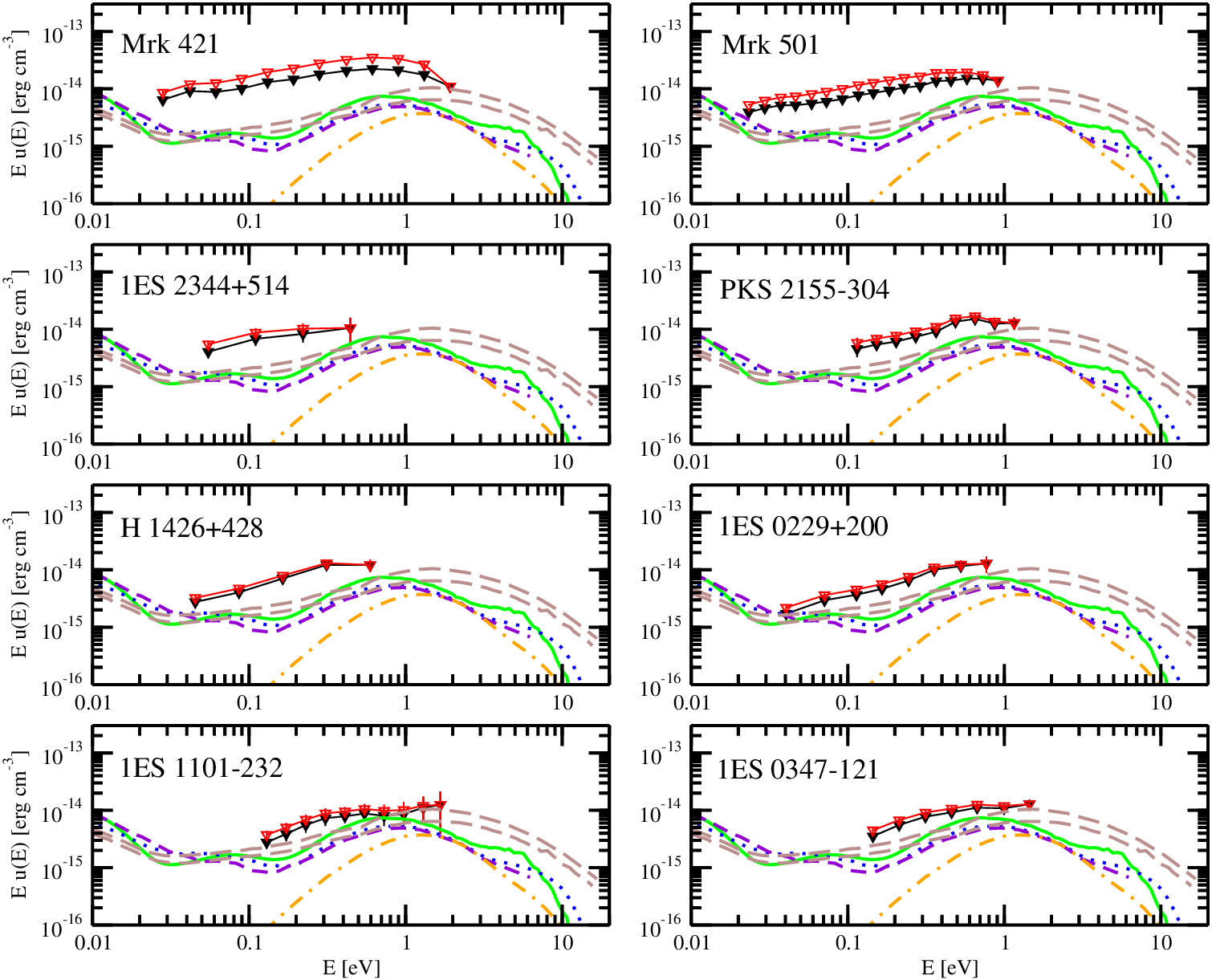}
\caption{ Upper limits for the conventional
($\Gamma_{int}^{min}=1.5$; black filled inverted triangles) and 
extreme ($\Gamma_{int}^{min}=1.0$; red empty inverted triangles)
limits on the EBL.  Also plotted are several EBL models: the best fit
model from \citet[][solid green curve]{kneiske04}, the fiducial model
from \citet[][dotted blue curve]{gilmore08}, the model of
\citet[][short dashed violet curve]{franceschini08}, the baseline
and fast evolution models of \citet[][lower and upper long dashed
brown curves, respectively]{stecker06}, and model B from
\citet[][dot-dashed orange curve]{razzaque09}.  }

\label{blazar_limits}
\end{figure*}

\section{Results and Discussion}

The blazar sample used to constrain the EBL is seen in Table
\ref{blazartable}.  This is not a complete sample of TeV blazar
spectra\footnote{For complete, continually updated catalogs see the
TeVCat (\url{http://tevcat.uchicago.edu/}) or R. Wagner's website
(\url{http://www.mppmu.mpg.de/~rwagner/sources/}).}, as we have
selected blazars that are nearby and should give the strongest
constraints, i.e.,that have the smallest $\xi$.  The EBL upper limits
from the 8 blazar spectra with the lowest $\xi$ are plotted in
Fig. \ref{blazar_limits}.

\begin{figure*}
\epsscale{1.0}
\plotone{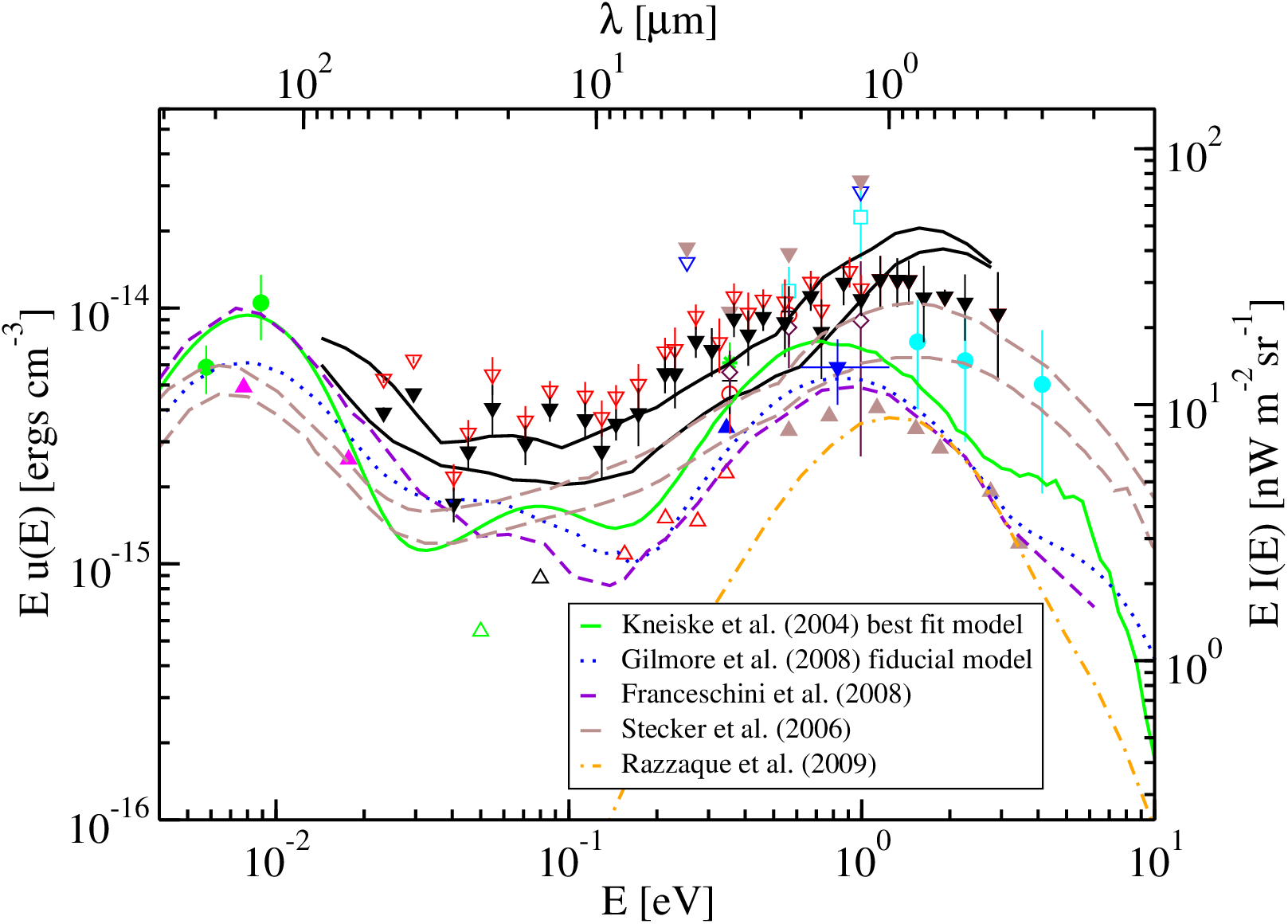}
\caption{
Measurements and constraints on the EBL energy density from observations.  
Measurements are from \citet[][cyan points]{bernstein02}, 
\citet[][empty red circle]{gorjian00}, 
\citet[][green asterisk]{dwek98_obs}, 
\citet[][empty cyan square]{cambresy01}, 
\citet[][black cross]{wright00}, 
\citet[][maroon diamonds]{levenson07}, and
\citet[][green filled circles]{hauser98}.  
Lower limits are from \citet[][red empty triangles]{fazio04}, 
\citet[][brown filled triangles]{madau00}, 
\citet[][blue filled triangle]{levenson08}, 
\citet[][magenta filled triangles]{dole06}, 
\citet[][black empty triangle]{metcalfe03}, and 
\citet[][green empty triangle]{papovich04}.  
Upper limits are from \citet[][brown filled inverted triangles]{hauser98}, 
\citet[][blue empty inverted triangles]{dwek98_obs}, 
\citet[][blue filled inverted triangle]{aharonian06},  
\citet[][upper and lower black curves $\Gamma_{int}^{min}=0.75$ and 
$\Gamma_{int}^{min}=1.5$ upper limits, respectively]{mazin07}, and 
Red empty and black filled inverted triangles are the  
$\Gamma_{int}^{min}=1.0$ and $\Gamma_{int}^{min}=1.5$ 
upper limits, respectively, from this paper.  The curves are the 
same models as in Fig. \ref{blazar_limits}.
}
\label{EBL_limits}
\end{figure*}

For each blazar, $i$, one gets a list of $\e u^{ max}_{EBL,j}(\e)$ at
each EBL energy bin $j$, which corresponds to an observed $\g$-ray
energy bin.  For $N$ such blazars, we choose the $\min\{\e u^{
max}_{EBL,j}(\e)\}$ to give the strongest constraint on the $j$th
energy bin.  These constraints, along with other EBL measurements and
constraints, are plotted in Fig. \ref{EBL_limits}.

In section \S\ \ref{EBLconstraint} we have made several
assumptions to get the upper limit, namely, that the EBL does not
evolve at low redshift, and the absorption occurs entirely from an EBL
photon of energy equal to the energy where the absorption
cross-section is maximized.  There is no direct way to measure the EBL
at higher redshifts, however the star formation rate increases between
$z=0.0$ and $z=0.2$ \citep[e.g.,][]{hopkins06} as does the luminosity
density at various wavelengths \citep[e.g.,][]{lefloch05, babbedge06}.
Thus it seems logical that the EBL energy density would be greater at
higher redshifts.  A greater EBL energy density in the past would
result in a greater absorption, which would give weaker upper limits
than what the actual limits would be if evolution was taken into
account.  Thus, if we did take evolution into account, our limits
would be stronger.  But the largest source of error is likely to be
the monochromatic assumption.  If the EBL energy density is
significantly greater at higher energies (shorter wavelengths) than
where the cross section is maximized ($\e_1\e^\prime = 2$), the result
is a considerably weaker upper limit.  In all of the EBL models
presented in Fig.\ \ref{EBL_limits} the EBL is considerably higher at
$\ga0.2$ eV ($\la 5\ \mu$m) than immediately below this energy.  Thus,
the EBL energy density at energies less than 0.2 eV is likely to be
considerably lower than our upper limits, although to actually
quantify this error one would need to assume an EBL model.  Since our
conservative assumptions all result in upper limits erring on the high
side, our constraints are quite strong.  

These EBL upper limits may be compared to the constraints by
\citet{mazin07}.  Their technique gives similar results and is in some
ways complementary to our results.  We use more constraining
blazars than \citet{mazin07}, particularly the blazar \object{1ES
0229+200}, which gives a stronger constraint at  $\sim 0.04$
eV ($\sim 30\ \mu$m).  The reported spectrum of \object{1ES
1101-232} \citep{aharonian06}, which was used by \citet{mazin07}, has
been re-analyzed with improved energy calibration
\citep{aharonian07_1101}, which we use instead.  This is one reason
our constraints are weaker than \citet{mazin07} in the 0.2--0.5 eV
(3--7 $\mu$m) range.  \citet{mazin07} also find that the size of the
EBL grid to which they fit splines is a cause of systematic error.  By
using a finer grid, they find they can get constraints 20--30\%
higher.  This could be another source of discrepancy between their
results and ours, since our results do not make any assumptions as to
the coarseness of the possible shape of the EBL spectrum.
Additionally, our monochromatic assumption could lead to a higher
upper limit, as discussed above.  \citet{mazin07} estimate their
errors in the optical and near-IR to be about 30\%.  Our results are
within these errors at 0.2--0.5 eV.  Our results are lower in
the optical ($\la 10$ $\mu$m or $\ga$ 1 eV) due to the assumption of
the highest possible EBL energy density.  They choose an arbitrary
level larger than we do, since we choose the largest possible model,
the fast evolution model of \citet{stecker06}.  The choice in this
region does not act as a constraint, since our results will naturally
be above the highest EBL model.  

Our results are consistent with the recent highly constraining EBL
lower limit at 0.34 eV (3.6 $\mu$m) \citep{levenson08} from number
counts with the {\em Spitzer Space Telescope} \citep[see][for a
different interpretation]{krennrich08}.  The lower limit of
\citet{levenson08} combined with our results and other data means that
the EBL is quite well constrained at 3.6 $\mu$m.  The only model above
this limit is the best fit model of \citet{kneiske04}.  However, their
model falls just above our conventional upper limit
($\Gamma_{int}^{min}=1.5$) at $\approx0.7$ eV ($\approx1.7\ \mu$m).
The models of \citet{gilmore08}, \citet{franceschini08}, and the
fast evolution model of \citet{stecker06} also fall above our 
conventional limit at $\approx 0.04$ eV ($\approx 30\ \mu$m), while
the baseline model of \citet{stecker06} is just below our upper limit.
None of the models are above the extreme limit
($\Gamma_{int}^{min}=1$), which is not surprising, considering this
weaker limitation was designed to avoid conflicts between $\g$-ray
observations and EBL models.  Future observations of blazars with ACTs
as well as the recently launched {\em Fermi Gamma-Ray Space Telescope}
will provide further upper limits on EBL models.

We have updated the technique of \citet{schroedter05_EBL} to a
recent sample of VHE $\g$-ray blazar spectra to obtain upper limits on
the EBL energy density.  This technique does not make assumptions
about the shape of the EBL or fit a certain model to a de-absorbed
spectrum, thus avoiding problems with assuming an EBL spectral shape
in other techniques for using VHE $\g$-ray observations to constrain
the EBL.  We obtain upper limits weaker and more conservative than
limits from de-absorbing spectra with an assumed EBL shape and fitting
the results with a power-law \citep[e.g.,][]{aharonian06} and in
agreement with a similar technique by \citet{mazin07}.  Our
conservative assumptions imply that it is likely the actual EBL energy
density is considerably lower than our upper limits, especially for
$m_ec^2\e \la 0.2$ eV.  

\acknowledgements 
We thank the anonymous referee and C. Dermer for useful comments
which have improved this paper.  S.R. is supported by the National
Research Council associateship program at the Naval Research
Laboratory.  J.D.F. was supported by NASA Swift Guest Investigator
Grant DPR-NNG05ED411 and NASA GLAST Science Investigation
DPR-S-1563-Y.



\bibliographystyle{apj}
\bibliography{references,blazar_ref,EBL_ref}

\begin{deluxetable}{lcccccl}
\tabletypesize{\scriptsize}
\tablecaption{
TeV Blazar Sample
}
\tablewidth{0pt}
\tablehead{
\colhead{ Blazar } &
\colhead{ Redshift } &
\colhead{ Spectral Index ($\Gamma$) } &
\colhead{ $E_{min}$ [TeV] } &
\colhead{ $E_{max}$ [TeV] } &
\colhead{ $\xi(\Gamma_{int}^{min}=1.5)$ } & 
\colhead{ Reference } 
}
\startdata
\object{Mrk 421} & 0.030 & $2.14\pm 0.10$ & 0.3 & 18 & 5.7 &\citet{krennrich01} \\
\object{Mrk 421} & 0.030 & $2.09\pm 0.09$ & 0.4 & 8 & 8.6 &\citet{krennrich02} \\
\object{Mrk 501} & 0.034 & $1.92\pm 0.15$ & 0.6 & 22 & 2.6 &\citet{aharonian99_mrk501} \\
\object{1ES 2344+514} & 0.044 & $2.54\pm 0.18$ & 1.2 & 11 & 8.6 &\citet{schroedter05_2344} \\
\object{Mrk 180} & 0.045 & $3.3\pm 0.7$ & 0.2 & 4.2 & 29.0 & \citet{albert06_mrk180} \\
\object{1ES 1959+650} & 0.047 & $3.18\pm 0.17$ & 1.5 & 11 & 9.1 & \citet{aharonian03_1959} \\
\object{BL Lacertae} & 0.069 & $3.64 \pm 0.54$ & 0.2 & 0.7 & 55.5 &\citet{albert07_bllac} \\
\object{PKS 2005-489} & 0.071 & $4.0 \pm 0.4$ & 0.2 & 2.5 & 35.6 & \citet{aharonian05_2005} \\
\object{RGB 0152+017} & 0.080 & $2.95\pm0.41$ & 0.3 & 3.0 & 16.4 & \citet{aharonian08_0152} \\
\object{W Comae} & 0.102 & $3.81\pm0.35$ & 0.3 & 1.1 & 35.1 & \citet{acciari08_wcomae} \\
\object{PKS 2155-304} & 0.116 & $3.19\pm0.17$ & 0.2 & 5 & 9.4 & \citet{aharonian07_2155} \\
\object{PKS 2155-304} & 0.116 & $2.84\pm0.24$ & 0.4 & 4 & 8.5 & \citet{aharonian05_2155}  \\
\object{H 1426+428} & 0.129 & $2.60\pm0.60$ & 0.8 & 10 & 3.3 & \citet{aharonian03_1426}  \\
\object{1ES 0229+200} & 0.139 & $2.50\pm0.21$ & 0.6 & 12 & 2.5 & \citet{aharonian07_0229} \\
\object{H 2356-309} & 0.165 & $3.09\pm0.26$ & 0.2 & 1 & 15.5 & \citet{aharonian06_2356} \\
\object{1ES 1218+304} & 0.182 & $3.00\pm0.40$ & 0.08 & 0.7 & 14.7 & \citet{albert06_1218} \\
\object{1ES 1218+304} & 0.182 & $3.08\pm0.41$ & 0.2 & 1.4 & 12.1 & \citet{acciari09_1218} \\
\object{1ES 1101-232} & 0.186 & $2.94\pm0.20$ & 0.3 & 3.3 & 6.6 & \citet{aharonian07_1101} \\
\object{1ES 0347-121} & 0.188 & $3.10\pm0.25$ & 0.3 & 3.0 & 7.7 & \citet{aharonian07_0347} \\
\object{1ES 1011+496} & 0.212 & $4.0\pm0.5$ & 0.2 & 0.6 & 21.6 & \citet{albert07_1011} \\
\enddata
\label{blazartable}
\end{deluxetable}

\clearpage
\clearpage

\end{document}